\documentclass[12pt,prd,aps,amssymb,amsmath,tightenlines,showpacs,a4paper]{revtex4}
\def\bea{\begin{eqnarray}}
\def\eea{\end{eqnarray}}
\def\beax{\begin{eqnarray*}}
\def\eeax{\end{eqnarray*}}
\def\half{\frac{1}{2}}
\def\quarter{\frac{1}{4}}
\def\vf{\varphi}

\begin{document}
\title{Green Functions for the Wrong-Sign Quartic}

\author{H.~F.~Jones\email{h.f.jones@imperial.ac.uk}}

\affiliation{Physics Department, Imperial College, London SW7 2BZ, UK}
\date{\today}

\begin{abstract}

It has been shown that the Schwinger-Dyson equations for non-Hermitian theories implicitly include the Hilbert-space metric. Approximate Green functions for such theories may thus be obtained, without having to evaluate the metric explicitly, by truncation of the equations. Such a calculation has recently been carried out for various $PT$-symmetric theories, in both quantum mechanics and quantum field theory, including the wrong-sign quartic oscillator. For this particular theory the metric is known in closed form, making possible an independent check of these approximate results. We do so by numerically evaluating the ground-state wave-function for the equivalent Hermitian Hamiltonian and using this wave-function, in conjunction with the metric operator, to calculate the one- and two-point Green functions. We find that the Green functions evaluated by lowest-order truncation of the Schwinger-Dyson equations are already accurate at the (6-8)\% level. This provides a strong justification for the method and a motivation for its extension to higher order and to higher dimensions, where the calculation of the metric is extremely difficult.

\end{abstract}

\pacs{02.30.Cj, 02.30.Mv, 03.65.Ca, 11.30.Er}
\maketitle

\section{Introduction}
The wrong-sign quartic oscillator, as we shall explain below, is a particularly interesting example of
a quantum theory whose Hamiltonian is not Hermitian, but is instead $PT$-symmetric, and as a consequence
of that unbroken $PT$ symmetry possesses a completely real energy spectrum. The recent surge of interest in this type of theory began with the pioneering paper of Bender and Boettcher\cite{BB}, in which they showed,
by numerical and asymptotic analysis, that the entire class of Hamiltonians
\bea
H=p^2-(ix)^M
\eea
had that property for $M\ge 2$. Apart from the trivial case $M=2$, the simplest example is for $M=3$, where
$H=p^2 +ix^3$, for which the Schr\"odinger differential equation can be analyzed on the real $x$ axis.
However, for $M\ge 4$ the eigenvalue equation must instead be formulated in the complex plane, inside
a Stokes wedge, in order that the wave-function vanish exponentially at infinity. In particular, for $M=4$ we obtain the wrong-sign quartic, $H=p^2-x^4$, which is unbounded below on the real axis, but can be understood
as a legitimate Hamiltonian when formulated on an appropriate contour in the complex $x$ plane.

Since that initial paper, there has been intensive investigation into the properties of Hamiltonians of this kind, whose progress can be followed in the reviews by Bender\cite{CMBR} and Mostafazadeh\cite{AMR}. We  restrict ourselves here to those features that form the essential background to the present calculation.

For a viable framework of quantum mechanics one needs not only a real spectrum but also a probabilistic interpretation. In standard quantum mechanics that is provided by matrix elements of the type $\int \psi^* \hat{A} \phi$, orthogonality of eigenfunctions
$\int dx\ \psi^*_1\psi_2=0$,  and the probability density $\psi^*\psi$. In $PT$-symmetric quantum mechanics it is found instead that orthogonality of eigenfunctions takes the nonlocal form
$\int dx\ \psi^*_1(-x)\psi_2(x)=0$, that is, $\int dx\ (\psi_1)_{PT}\psi_2=0$. This is a problem
because the metric involved in the corresponding normalization integral $\int dx\ (\psi)_{PT}\psi$ is not positive-definite, and we do not in the first instance have a proper probabilistic interpretation.
However, it was subsequently found\cite{BBJC} that another metric could be constructed, with the help of a grading operator $C$, which preserved orthogonality and gave a positive normalization integral $\int dx\ (\psi)_{CPT}\psi$. In contrast to standard quantum mechanics, this metric is not universal, but is dynamically determined by the particular Hamiltonian in question. The calculation of this metric is
extremely difficult, and in most cases can only be calculated approximately, either through a set of algebraic relations\cite{BBJQ}, or through the use of Moyal brackets\cite{SGF}.

A more general framework, of which $CPT$-symmetry is a special case, was developed by Mostafazadeh\cite{AMh}.
A Hamiltonian $H$ is said to be quasi-Hermitian if it can be related to a Hermitian Hamiltonian $h$
by a similarity transformation:
\bea
H=\rho^{-1} h \rho,
\eea
where $\rho$ is a positive-definite Hermitian operator. From this we immediately see that
\bea
H^\dag=\eta H \eta^{-1},
\eea
where $\eta=\rho^2$.
The connection with the $CPT$ formulation is that $\eta$ can be identified as $e^{-Q}$ when $CP$
is written\cite{BBJQ} in the exponential form $CP=e^Q$.
Accordingly $\rho$ can be written as $\rho=e^{-\half Q}$.

Matrix elements of operators should always be of the form $\langle \psi_1|\eta A |\psi_2\rangle$, i.e. calculated with inclusion of the metric. This result is easily obtained by a similarity transformation from the corresponding matrix element in the Hermitian theory. Thus, if $|\vf_{1,2}\rangle$ are the corresponding
states and $a$ the corresponding operator,
\bea\label{simy}
\psi_i=e^{\half Q} \vf_i, \hspace{1cm}
A=e^{\half Q}a e^{-\half Q}.
\eea
Hence\cite{fn1}
\bea
\langle \vf_1|a|\vf_2\rangle
=\langle \psi_1 |e^{-\half Q}(e^{-\half Q} A e^{\half Q})e^{-\half Q}|\psi_2\rangle
=\langle \psi_1 |e^{-Q} A |\psi_2\rangle.
\eea
In view of Eq.~(\ref{simy}), we can also write
\bea\label{GF}
\langle \psi_1 |e^{-Q} A |\psi_2\rangle= \langle \vf_1|e^{-\half Q} A e^{\half Q}|\vf_2\rangle,
\eea
a relation we shall need below.

When one is calculating Green functions using path integrals, the Schwinger-Dyson equations or Feynman diagrams, where the metric makes no explicit appearance, it is not immediately clear that the quantities so calculated correspond to matrix elements evaluated with the inclusion of the metric. This issue was addressed in Ref.~\cite{JR}, with the conclusion that this is indeed the case, the essential point being that only when the metric is correctly included do the Heisenberg equations of motion and the canonical commutation relations take their standard form. This leaves the way open to using non-perturbative truncations of the
Schwinger-Dyson equations to obtain approximate values for the Green functions of a quasi-Hermitian theory without the need to evaluate the metric, which, as mentioned above, is generally extremely difficult to calculate.

Such a non-perturbative calculation has recently been carried out by Bender\cite{SDEqs}, for various quasi-Hermitian quantum theories, and has the potential to be extended to quantum field theories, where the calculation of the metric is even more problematic. The method involves truncation of the Schwinger-Dyson by setting to zero the connected Green functions beyond a certain order, and has been applied, among other theories, to both the standard anharmonic oscillator and the wrong-sign anharmonic oscillator with or without a quadratic term. This last theory is special among quasi-Hermitian theories in that it is essentially the only non-trivial theory where the metric is known exactly\cite{JM}. Hence in this particular case we have the possibility of calculating the Green functions numerically, by utilizing the similarity transformation to the equivalent Hermitian theory, thus providing a check of the accuracy of the Schwinger-Dyson calculation.

The paper is organized as follows. In the next section we summarize, for completeness, the method of Ref.~\cite{SDEqs} and the results obtained there, specifically the renormalized mass and the first two connected Green functions $G_1$ and $G_2$ for the wrong-sign quartic oscillator without a quadratic term.
In this case we have scaling, and the result is universal. One can also easily include a quadratic term, but then one has a one-parameter set of solutions depending on the quadratic coefficient. In the following section we show why previous calculations of the Green functions, which predated the realization that matrix elements should be calculated with the inclusion of the metric, are in fact correct. We then describe how to calculate the Green functions using the Schr\"odinger wave-functions of the equivalent Hermitian Hamiltonian $h$, finding that, even in the simplest truncation, the Schwinger-Dyson equations give quite reasonable values. We conclude with a brief discussion of the significance of these results.

\section{Truncated Schwinger-Dyson Equations}
In this section we summarize the results obtained in Ref.~\cite{SDEqs}, which to this order essentially amount to mean field theory. The starting point is the
expression for the vacuum-generating functional as a Euclidean functional integral\cite{D} :
\bea
Z[J]=\int [d\vf]\ \exp\left[-\int d x \left(\half(\vf')^2+\half m^2\vf^2-\quarter g \vf^4 -J\vf\right) \right].
\eea
The connected one- and two-point functions, in the presence of $J$, are respectively defined as
\bea
G_1(x)&=&\frac{\delta}{\delta J(x)} \ln Z[J]\label{G1SD}\\
G_2(x,y)&=& \frac{\delta}{\delta J(y)}G_1(x) .
\eea
The Schwinger-Dyson equations are derived from the Heisenberg equation of motion for $\vf$:
\bea
(-\partial^2 \vf(x)- m^2) \vf -g\vf^3(x)=J(x) .
\eea
This becomes an equation for Green functions when we take its expectation value in the vacuum,
still with $J\ne 0$:
\bea\label{phi3}
(-\partial^2-m^2) G_1(x) -g\frac{\langle\langle 0|\vf^3(x)|0\rangle\rangle_J}{\langle\langle 0|0\rangle\rangle_J}=J(x).
\eea
Here the $\langle\langle\dots \rangle\rangle$ notation means that the expectation value is to be taken with the appropriate
metric, but the precise form of that metric plays no part in the equations that follow. $Z$ itself is
given by $Z[J]=\langle\langle 0|0\rangle\rangle_J$.

Taking successive functional derivatives $\delta/\delta J(x)$ of the identity $ZG_1(x)=\langle\langle 0|\vf(x)|0\rangle\rangle_J$ of Eq.~(\ref{G1SD}) we obtain
\bea
ZG_1^3(x)+3ZG_1(x) G_2(x,x)+ZG_3(x,x,x)=\langle\langle 0|\vf^3(x)|0\rangle_J ,
\eea
which allows us to substitute for $\langle\langle 0|\vf^3(x)|0\rangle_J$ in Eq.~(\ref{phi3}), to obtain
\bea\label{m1}
(-\partial^2-m^2) G_1(x) -g\left[G_1^3(x)+3G_1(x)G_2(x,x)+G_3(x,x,x)\right]=J(x).
\eea
Differentiating once more, $\delta/\delta J(y)$, gives the equation
\bea\label{m2}
(-\partial^2-m^2) G_2(x) -g\left[3 G_1^2(x)G_2(x,y)+3G_2(x,y)G_2(x,x)\right.&&\nonumber\\
\left.+3G_1(x)G_3(x,x,y)+G_4(x,x,x,x)\right]&&=\delta(x-y).
\eea
We now set $J=0$, in which case $G_1(x)$ becomes a constant independent of $x$, and $G_2(x,y)$ becomes
a function of $x-y$, so in a change of notation we write it as a function of that single argument: $G_2(x-y)$.
Finally we truncate Eqs.~(\ref{m1}) and (\ref{m2}) by neglecting $G_3$ and $G_4$, to obtain
\bea
m^2 G_1 -g\left[G_1^3+3G_1 G_2(0)\right]&=&0\nonumber , \\
&&\\
\left[-\partial^2+m^2-3g\left(G_1^2+G_2(0)\right)\right]G_2(x-y)&=&\delta(x-y)\nonumber .
\eea
Assuming\cite{G1eq0} that $G_1\ne 0$, the first equation gives
\bea\label{first}
m^2-g\left(G_1^2+3G_2(0)\right)=0 ,
\eea
while the second equation gives
\bea
G_2(x)=\frac{e^{-M|x|}}{2M},
\eea
where
\bea
M^2=m^2-3g(G_1^2+G_2(0)),
\eea
so that $G_2(0)=1/(2M)$. Hence we have
\bea
M^2=-2g G_1^2= -2m^2+\frac{3g}{M}
\eea
Let us restrict ourselves to the case $m=0$, in which case\cite{fn}
\bea
M&=&(3g)^{1/3},\nonumber \\
iG_1&=&\frac{M}{(2g)^\half},\\
G_2(0)&=&\frac{1}{2M}\nonumber .
\eea
 Since in the massless case the results
scale with various powers of $g$, we may as well take some specific value. For ease of comparison
with the calculations of the following section we choose $g=2$, in which case
\bea\label{SDEst}
M&=& 6^{1/3}=1.81712\nonumber ,\\
iG_1&=&\half M=0.90856,\\
G_2(0)&=&\frac{1}{2M}=0.27516.\nonumber
\eea
Note that for $g=-2$, the conventional anharmonic oscillator, the corresponding results are
$M=3^{1/3}$, $G_1=0$, $G_2(0)=1/(2M)$.

\section{Alternative Methods for Calculating Green Functions}
In fact these Green functions have been estimated already, in a paper that preceded the introduction of the
$\eta$ metric\cite{BMY}, in two different ways.

The first was to calculate $\langle\langle z^n\rangle\rangle$ in terms of the Schr\"odinger wave-function as
\bea\label{BMY1}
\langle\langle z^n \rangle\rangle=\frac{\int_C dz \psi_0(z)^2 z^n }{\int_C dz \psi_0(z)^2 },
\eea
where we write $z$ instead of $x$ to emphasize that the theory must be formulated on an appropriate
contour $C$ in the complex $z$ plane. This is the first difference from a conventional Hermitian theory.
The second is that the expectation value is taken with $\psi_0(z)^2$ rather
than $|\psi_0(z)|^2$. This can be motivated by the fact that the former, as an analytic function, can be analytically continued, whereas the latter cannot, and moreover it is line with perturbation theory, which involves $\psi_0$ alone, not its complex conjugate.

However, in the light of what we have subsequently learnt about the metric of such theories, it is natural to ask whether Eq.~(\ref{BMY1}) does indeed correspond to the expectation value $\langle\langle z^n \rangle\rangle$,
which explicitly involves the metric. The answer is in the affirmative because of the following identity.

Recall that from Eq.~(\ref{simy}) $\psi_0=e^{\half Q}\vf_0$, where $\vf_0$, the ground-state wave-function of a Hermitian Hamiltonian, can be chosen to be real. Thus
\bea
\psi_0^*&=&e^{\half Q^*}\vf_0\nonumber\\
&=&e^{-Q}\psi_0,
\eea
since $Q(x,y)$ (see Eq.~(\ref{Q4})) is pure imaginary (and antisymmetric).
Thus, in Eq.~(\ref{BMY1}) we can take the first factor of $\psi_0$ as $\psi_0=e^Q\psi_0^*$. Then, under integration by parts, or equivalently using the antisymmetry of $Q$, we obtain
\bea
\int dz \psi_0 z^n \psi_0 = \int dz \psi_0^* e^{-Q} z^n \psi_0,
\eea
so that Eq.~(\ref{BMY1}) indeed gives the correct formula for $\langle\langle z^n \rangle\rangle$, taking into account the metric.

The second method used in Ref.~\cite{BMY}, which can be extended to higher dimensions, was to apply the strong-coupling approximation to the functional integral. Here again the metric does not enter explicitly, but nonetheless, as explained in Ref.~\cite{JR}, the
metric is implicitly involved, and the Green functions so calculated are of the type $\langle\langle z^n\rangle\rangle$.

\section{Green Functions from the Equivalent Hermitian Hamiltonian}
We start with
\bea
H_z=p_z^2-\lambda z^4,
\eea
where, as in the previous section, we write $z$ instead of $x$ to emphasize that the theory must be formulated in the complex $z$ plane.
The connection with the Hamiltonian of Section II is that $H_z=2 H_{SD}$ with $\lambda=\half g = 1$.

There is one particular contour, parametrized by the real variable $x$, which makes the subsequent
calculations practicable, namely\cite{JM}
\bea
z=-2\surd(1+ix).
\eea
This gives an intermediate non-Hermitian Hamiltonian
\bea
H_x=\half\{(1+ix), p_x^2\} -\half p_x -\alpha(1+ix)^2,
\eea
where $\alpha=16\lambda$.
For this Hamiltonian $Q$ can be found explicitly as
\bea\label{Q4}
Q=-\frac{p^3}{3\alpha} +2 p\  ,
\eea
and the equivalent Hermitian Hamiltonian $h$ as
\bea
h\equiv e^{-\half Q} H e^{\half Q}=\frac{p^4}{4\alpha}-\half p +\alpha x^2\ .
\eea
Since $h$ is quartic in $p$ but quadratic in $x$ it is natural to work with the Fourier-transformed wave-function $\tilde{\vf}(p)$.
The Schr\"odinger equation for $\tilde\vf(p)$ is
\bea
\left(-\alpha \frac{\partial^2}{\partial p^2}-\half p +\frac{p^4}{4\alpha}\right)\tilde\vf(p)=E\tilde\vf(p).
\eea
Scaling $p$ to $p=y\sqrt{\alpha}$ we get
\bea\label{scaledh}
\left(- \frac{\partial^2}{\partial y^2}-\half y \sqrt{\alpha}+\alpha\frac{y^4}{4}\right)\tilde\vf=E\tilde\vf.
\eea
Recall that we have chosen $\lambda=1$, so that $\alpha=16$.

Before going on to the Green functions let us first check $M$, which in terms of
quantum-mechanical energy levels is the gap between the first excited state and the ground state, divided
by two because $H_z=2 H_{SD}$, namely $M=\half(E_1-E_0)$.
The numerical eigenvalues obtained from Eq.~(\ref{scaledh}) are
\bea\label{E0andE1}
E_0&=&1.4771497535\nonumber\\
&&\\
E_1&=&6.0033860834\ ,\nonumber
\eea
in agreement with Bender and Boettcher\cite{BB}, who obtained these results by numerically solving
the eigenvalue equation along a complex contour. Eq.~(\ref{E0andE1}) gives $M=2.26312$, showing that the estimate 1.81712 from the lowest-order truncation of the Schwinger-Dyson equations is not particularly accurate.

Turning our attention now to the Green functions, we use the result (cf. Eq.~(\ref{GF})) that the expectation value of any operator $\cal O$ is obtained in terms of the ground-state WF $\tilde{\vf}_0(p)$ of $h$ as
\bea
\langle {\cal O} \rangle = \langle \tilde{\vf}_0(p)| e^{-\half Q}{\cal O} e^{\half Q}|\tilde{\vf}_0(p)\rangle\ .
\eea
So in particular
\bea\label{G1}
\langle z \rangle = -2i \langle \tilde{\vf}_0(p)| e^{-\half Q}(1+ix)^\half e^{\half Q}|\tilde{\vf}_0(p)\rangle\ ,
\eea
where $x=i\partial/\partial p$.

It is in fact easier to calculate
\bea\label{z2h}
\langle z^2 \rangle = -4 \langle \tilde{\vf}_0(p)| e^{-\half Q}(1+ix) e^{\half Q}|\tilde{\vf}_0(p)\rangle\ ,
\eea
whose S-D estimate is
\bea\label{z2SD}
\langle z^2 \rangle_{\rm SD} = G_2(0)+G_1^2= -0.5503\ .
\eea
We now check this from Eq.~(\ref{z2h}). Note that
\bea
ix e^{\half Q}\tilde\vf_0(p) &=& -\frac{\partial}{\partial p}\left(e^{\half Q}\tilde\vf_0(p)\right)\nonumber\\
&=&e^{\half Q}\left[-\left(1-\frac{p^2}{2\alpha}\right)-\frac{\partial}{\partial p}\right]\tilde\vf_0(p)
\eea
So
\bea
\langle z^2 \rangle = -\frac{1}{8 \lambda}\langle\tilde{\vf}_0(p)| p^2|\tilde{\vf}_0(p)\rangle ,
\eea
since $\tilde\vf_0(p)$ can be taken real, so that the derivative term vanishes on integration.
In terms of $y$, this is
\bea
\langle z^2 \rangle = -2\langle\tilde{\vf_0}| y^2|\tilde{\vf}_0\rangle .
\eea
The numerical value, found by evaluating this expectation value with the numerical ground-state
wave-function $\tilde{\vf(y)}$ is $\langle\tilde{\vf_0}| y^2|\tilde{\vf}_0\rangle=0.2585$.
Hence
\bea
\langle z^2 \rangle=-0.517,
\eea
which is actually quite close to the Schwinger-Dyson estimate -0.5503 of Eq.~(\ref{z2SD}).\\

Returning now to the calculation of $\langle z \rangle$ from (\ref{G1}), we deal with $(1+ix)^\half$ by writing it as\cite{RJR}
\bea\label{identity}
(1+ix)^\half&=&(1+ix)(1+ix)^{-\half}\nonumber\\
&=&(1+ix)\int_{-\infty}^\infty \frac{du}{\sqrt{\pi}}\ e^{-u^2(1+ix)}\nonumber\\
&=& \left(1-\frac{\partial}{\partial p}\right) \int_{-\infty}^\infty \frac{du}{\sqrt{\pi}}\ e^{-u^2}\ e^{u^2\partial/\partial p}
\eea
Then
\bea
\langle z \rangle &=& -\frac{2i}{\sqrt{\pi}} \int_{-\infty}^\infty dp\ \tilde{\vf}_0(p)\ du\ e^{-u^2} e^{-\half Q(p)} (1-\partial/\partial p)
e^{u^2\partial/\partial p}\left(e^{\half Q}\tilde{\vf}_0(p)\right)\nonumber\\
&=&-\frac{2i}{\sqrt{\pi}} \int_{-\infty}^\infty dp\ \tilde{\vf}_0(p)\ du\ e^{-u^2} e^{-\half Q(p)} (1-\partial/\partial p)
\left(e^{\half Q(p+u^2)}\tilde{\vf}_0(p+u^2)\right)\nonumber\\
&=&-\frac{2i}{\sqrt{\pi}} \int_{-\infty}^\infty dp\ \tilde{\vf}_0(p)\ du\ e^{-u^2}e^{\half(Q(p+u^2)-Q(p))}
\times\nonumber\\
&&\hspace{2cm}\times \left[\tilde{\vf}_0(p+u^2)\left(1-\half Q'(p+u^2)\right)- \tilde{\vf}'_0 (p+u^2)\right]\nonumber\\
&=&-\frac{2i}{\sqrt{\pi}} \int_{-\infty}^\infty dp\ \tilde{\vf}_0(p)\ du\ e^{-u^2} e^{\half(Q(p+u^2)-Q(p))}
\times\\
&&\hspace{2cm}\times \left[\frac{1}{2\alpha}(p+u^2)^2\tilde{\vf}_0(p+u^2)- \tilde{\vf}'_0 (p+u^2)\right]
\nonumber
\eea
Changing variables according to $u^2=v\sqrt{\alpha}$ and $p=y\sqrt{\alpha}$, this becomes
\bea
\langle z \rangle &=&-\frac{2i\alpha^{\quarter}}{\sqrt{\pi}} \int_{-\infty}^\infty dy\ \tilde{\vf}_0(y)\ \int_0^\infty  \frac{dv}{\sqrt{v}}\
e^{-v \sqrt{\alpha}} e^{\half(Q(y+v)-Q(y))}
\times\nonumber\\
&&\hspace{2cm}\times \left[\half(y+v)^2\tilde{\vf}_0(y+v)- \frac{1}{\sqrt{\alpha}}\ \tilde{\vf}'_0 (y+v)\right]\ ,
\eea
where $Q(y)=\sqrt{\alpha}\ (2y -y^3/3)$.

By numerical integration we obtain
\bea
\langle z \rangle=-i \times 0.869\ ,
\eea
compared with the Schwinger-Dyson estimate $-i \times 0.908$ of Eq.~(\ref{SDEst}). We should also note that, when scaled to the value $g=1$, our results agree with those of Ref.~\cite{BMY}, apart from the fact that the sign of $G_2$ should be positive, not negative as it is there.

\section{Discussion}

We have calculated the Green functions for the wrong-sign quartic by exploiting the similarity transformation to an equivalent Hermitian Hamiltonian, and used these as a check of the method of truncation of the Schwinger-Dyson equations. We explicitly discussed the case $m=0$, but the extension to $m\ne 0$ is straightforward, as is  the extension to higher-order Green functions once the identity (\ref{identity}) is taken into account. We have also explained why previous calculations of these Green functions, which did not explicitly involve the metric, nonetheless incorporate it correctly.

In quantum mechanics it happens that we are able to calculate the metric exactly, but the method unfortunately does not extend to higher dimensions. However, for the purposes of evaluating Green functions it is not necessary because one can use non-perturbative techniques applied to the Schwinger-Dyson equations or the associated functional integrals, in the knowledge that such methods are guaranteed to evaluate the correct Green functions.
As far as the functional integral is concerned, the strong-coupling expansion used in Ref.~\cite{BMY} is a viable method only for the $m=0$ case. Unfortunately, Monte-Carlo simulation is not viable, because of the complex nature of the functional integral.  Altogether it seems that the Schwinger-Dyson approach is the most promising one.


\begin{thebibliography}{99}
\bibitem{fn1} Note that if $Q$ had been anti-Hermitian, giving a unitary transformation instead
of a similarity transformation, the first factor of $e^{-\half Q}$ would have become $e^{+\half Q}$,
thus cancelling with the second, and leaving no metric factor.
\bibitem{D} This functional integral is written for a field theory in an arbitrary number of spatial
 dimensions $D$. For quantum mechanics $D=1$, and $\vf(x)$ is really $x(t)$.
\bibitem{G1eq0}If $G_1=0$ the equation for $M$ has only a physically unacceptable negative
 solution.
\bibitem{fn}The sign of $G_1$ is not determined by these equations, but we have chosen it so as to agree with the
calculations of Sections III and IV.
\end{thebibliography}
\end{document}